\documentstyle[epsfig,here,12pt]{article}
\tiny\normalsize

\begin{document}
\def\MSbar {\hbox{$\overline{\hbox{MS}}\,$}}
\def\beq{\begin{equation}}
\def\eeq{\end{equation}}
\def\MSbar {\hbox{$\overline{\hbox{MS}}\,$}}

\vskip 20pt

\begin{titlepage}
\begin{flushright}
{\footnotesize
TAUP-2451-97\\}
\end{flushright}
\begin{centering}
{\Large{\bf
Resummation of perturbative QCD by Pad\'e approximants}}\footnote{
A lecture given at the Cracow School of Theoretical Physics,
Zakopane, May 30 -- June 10, 1997}\\
\vspace{2cm}
{\bf Einan Gardi} \,\,\, \\
\vspace{.4cm}
School of Physics and Astronomy
\\ Raymond and Beverly Sackler Faculty of Exact Sciences
\\ Tel-Aviv University, 69978 Tel-Aviv, Israel
\\ e-mail: gardi@post.tau.ac.il
\\
\vspace{2cm}

{\bf Abstract} \\
\vspace{0.35cm}
{\small
In this lecture I present some of the new developments concerning
the use of Pad\'e Approximants (PA's) for resumming perturbative series in QCD.
It is shown that PA's tend to reduce the renormalization scale and
scheme dependence as compared to truncated series.  
In particular it is proven that in the limit where the $\beta$
function is dominated by the 1-loop contribution, there is an exact symmetry
that guarantees  invariance of diagonal PA's under changing the 
renormalization scale.
In addition it is shown that in the large $\beta_0$ approximation 
diagonal PA's can be interpreted as a systematic 
method for approximating the flow of momentum in Feynman diagrams.
This corresponds to a new multiple scale generalization of the 
Brodsky-Lepage-Mackenzie (BLM) method  to higher orders.
I illustrate the method with the Bjorken sum rule and the vacuum
polarization function.
} 
\end{centering}

\end{titlepage}
\vfill\eject
\vskip 15pt
I will talk about resummation of perturbative series in QCD
\footnote{For a recent review on
the nature of perturbative series and resummation techniques 
see ref.~\cite{Fischer}}. The basic
question I deal with is how to use finite order perturbative
calculations in QCD to make unambiguous theoretical predictions, with
controlled errors. As experiments improve one requires the
theoretical predictions to be more accurate. However, in QCD it is
very hard to get accurate predictions, basically because the coupling
constant is large. This leads to non-negligible non-perturbative
effects as well as a badly divergent and renormalization scheme
dependent perturbative series.

In this talk I will show that PA's which start out as an
alternative to a finite order perturbative series having the same
formal accuracy, actually have an
important advantage over the finite order series\footnote{PA have
  various successful application in physics. Examples of applications
  to statistical physics and quantum field theory are listed in
  ref.~\cite{padeworks}. Applications to QCD appear in ref.~\cite{PA_QCD}}. 
Through the resummation of certain all-order effects related 
to the running of the coupling-constant, PA's become independent of
the choice of the renormalization scale and therefore lead to more
accurate and more reliable predictions. The material presented in this
lecture appears in greater detail in ref.~\cite{PA_QCD,Why,MDF}.

The outline of the talk is as follows: I will start by introducing
the PA's method and the problem of renormalization scale
dependence in QCD. I will mention some of the other ideas that were
raised to confront the problem of renormalization scale 
dependence and show how PA's solve it in a most elegant way.
Then I will address the question of what higher order effects are
summed-up by PA. I will show that there is a direct
interpretation of PA's in terms of approximating the momentum distribution
of virtual gluons in Feynman diagrams.

First, what are PA and how do I use them?
I start with an effective charge related to some physical observable,
written as a power series in $x$:
\begin{equation}
S_n = x\left(1\,+\,r_1 x \,+\,r_2 x^2\, +\,\cdots \,+\,r_n x^n\right)
\label{S_n}
\end{equation}
where $x=\alpha_s/\pi$. A PA is constructed by
writing a ratio of two polynomials such that when expanded back to a
Taylor series, it gives the known coefficients $r_1$ through $r_n$:
\[
P_{x[N/M]} = x \, \frac{1 + a_1x + ... +a_{N}x^N}{1 + b_1x + ... + b_Mx^M}\,\,\,\,\,\,\,\,\,\,\,\,\,\,\,\,,\,\,\,\,N+M=n
\]
There is a theorem that for any degree $N$ in the numerator and $M$ in
the denominator such that $N+M=n$ there is a unique PA function \cite{Baker}.
I will mainly deal here with diagonal PA's which are written, 
in my notation,  
as $x[N/N+1]$, having one power of $x$ out of the brackets.
I will soon come back to discuss PA's.

A renormalized perturbative series in QCD is not expected to give
exact predictions for measurable
quantities due a few limitations. First, the series is divergent and
not even Borel-summable. The resulting ambiguity is 
related to the  existence of non-perturbative effects. 
Second, at any given order, the partial-sum
depends on non-physical parameters, such as the renormalization
scale. This also makes the prediction ambiguous. 

Let us concentrate on the renormalization scale dependence.
Usually, when we calculate some observable $R$ that depends on one
external momentum $Q^2$ in perturbation theory, we choose as an
expansion parameter the renormalized coupling-constant at the
external scale $Q^2$. This ``natural'' choice of $\mu^2=Q^2$ is, 
however, quite arbitrary. We can, just as well, use some
other expansion parameter $y=\alpha_s(e^t Q^2)/\pi$, where $t\neq0$.
The renormalization group equation
\[
\frac{dx}{dt} \,=\, \beta_0x^2\,+\,\beta_1x^3\,+\,\beta_2x^4+\cdots 
\]
determines how the two couplants are related, 
\begin{equation}
x = y+ \beta_0 t \,y^2 +  \left(\beta_0^2 t^2 + \beta_1 t\right) \,y^3 + \left(
\beta_0^3 t^3 + \frac{5}{2} \beta_1 \beta_0 t^2 + \beta_2 t \right)\, y^4 +
 \cdots
\label{xy}
\end{equation}
and thus how the finite order series can be written in terms
of $y$:
\begin{equation}
\tilde{S}_n(t)=y\, \left(\,1\,+\,\tilde{r}_1\,y\,+\, \tilde{r}_2\,y^2\,+\,
  \tilde{r}_3\,y^3\,... +\, \tilde{r}_n\,y^n\,\right). 
\label{tilde_S_n}
\end{equation}
 The new coefficients $\tilde{r}_i$ are different from the original 
coefficients $r_i$, so as to compensate for the scale
shift, such that the total effect is some residual dependence on
$t$ which is of the next, uncalculated order. 
Still, in QCD, since the coupling constant is
large, the numerical difference due to the change of scale can be
quite large.  This limits the predictive power of the theory.
Beyond two-loops, there is also the question of scheme dependence
which can be parameterized by the higher-order coefficients of the $\beta$
function, $\beta_2$, $\beta_3$ and onward.

In order to test the significance of this scale and scheme dependence,
we studied \cite{PA_QCD} the polarized Bjorken Sum-Rule.  
In fig.~1 the Bjorken effective charge at NNLO for $Q^2\, =\, 20\,GeV^2$  
is plotted
as a function of the renormalization group non-physical parameters: 
the coupling $x=\alpha_s(\mu^2)/\pi$ and the
second coefficient of the $\beta$ function: $C_2=\beta_2/\beta_0$. 
We see that the surface is far from being flat.

The same surface is drawn again in fig.~2, but here -- as a
contour plot. The thick lines are contours of equal effective
charge. 
Large renormalization scheme dependence corresponds to large
higher-order corrections, since these are required to compensate for
the scale dependence. This observation makes it clear that we
should carefully choose the renormalization scale and scheme that we
are using.  
In fig.~2 one can identify a region of relatively low 
renormalization scale and scheme dependence. Specific scales and
schemes are
chosen according to different criterions such as the method of
Effective Charges \cite{ECH}, the Principal of Minimal
Sensitivity \cite{PMS} and the BLM scale-setting method \cite{BLM}. 
For the Bjorken Sum-Rule example (fig.~2), all of the above are 
located in the central region of low renormalization scale and scheme
dependence. Note that in this case,
 $\MSbar$, with $\mu^2=Q^2$ is not a good choice.

Let's go back to eq.~(\ref{xy}) that describes the scale transformation 
relating the coupling-constants $x$ and $y$ defined at two different scales.
If we assume that the 1-loop
coefficient of the $\beta$ function, $\beta_0$, is large enough, i.e.
\[
\, \beta_0 \gg \beta_i x^i
\]
for any $i \geq 1$, we can approximate the full relation by one that 
includes only the leading terms in $\beta_0$:
\[
x \simeq y+ \beta_0 t \,y^2 +  \beta_0^2 t^2 \,y^3 + 
\beta_0^3 t^3 \, y^4 + \cdots
\]
This can be written in a closed form:
\[
x = \frac{y}{1-\beta_0 t y}.
\]

It is important to realize that in the physical case of QCD with 
3 to 5 flavors, this
approximation is good. Fig.~3 shows the renormalization scale 
transformation itself, namely the
running coupling constant as a function of the scale. 
The dashed  line is
the best we know of the running coupling in QCD (it includes the 4-loop
effects), and the solid line is the 1-loop running coupling.
The two are quite close and I shall use here the 1-loop formula.

I now get to the main point, which is the independence of PA on the
renormalization scale.
We saw that partial-sums as usually written in perturbation theory
always yield different results in different renormalization scales:
\[
S_n(0) \,\neq\,\tilde{S}_n(t)
\]
where $S_n(0)$ refers to $\mu^2=Q^2$ as in eq.~(\ref{S_n}), 
$\tilde{S}_n(t)$ refers to $\mu^2=e^t Q^2$ as in 
 eq.~(\ref{tilde_S_n}), and  $x = {y}/\left({1-\beta_0 t y}\right)$.
However, 
if we construct a diagonal PA from the series in $x$ (eq.~(\ref{S_n})),
\[
P_{x[N/N+1] }(x)\,=\, x \frac{1 + a_1x + ... +a_{N}x^N}
{1 + b_1x + ... + b_{N+1}x^{N+1}}
\]
 and
independently, another PA from the series in $y$ (eq.~(\ref{tilde_S_n})), 
\[
\tilde{P}_{y[N/N+1] }(y)\,=\, y \frac{1 + \tilde{a}_1y + ... +
\tilde{a}_{N}y^N}
{1 + \tilde{b}_1y + ... + \tilde{b}_{N+1}y^{N+1}}
\]
we will get the same result in both: 
\[
P_{x[N/N+1] }(x)\,=\,\tilde{P}_{y[N/N+1] }(y).
\]
This is due to the mathematical property of
diagonal PA's: they are invariant under homographic transformations 
of the PA argument $(x \longrightarrow x/(1+Kx)$, see \cite{Why,Baker}).
We know that the all-order result does not depend on the
renormalization scale. The fact that diagonal PA are invariant
suggests that they correctly resum certain all-order effects that are
related to the running of the coupling-constant.

Non-diagonal PA are not exactly invariant. However, on the global
level (for large scale shifts $t$) they always have a reduced scale
dependence  \cite{Why}.
Going back to the example we examined above, namely the NNLO 
Bjorken sum-rule, we show in fig.~4 the $x[0/2]$ PA. Clearly (compare
with the partial-sum of fig.~1) the
renormalization scale and scheme dependence is almost 
completely eliminated! 

Non-diagonal PA's may be dangerous, since specific renormalization
scales and schemes are sometimes particularly deviant, as in the example of
the $x[1/1]$ PA for the Bjorken sum-rule shown in fig.~5. 
Therefore it is best to use a diagonal $x[N-1/N]$ PA.

I now consider the question of what higher-order contributions are
summed-up by diagonal PA's. It turns out that we can get some rigorous
results \cite{MDF} if we limit ourselves to the ``large $\beta_0$''
approximation \cite{mom,BB,LTM}, 
where only the leading term in $\beta_0$ in each
perturbative coefficient is taken into account. 
This approximation corresponds to summing certain higher-order 
contributions that are related to the
exchange of {\em one} virtual gluon. 
I use Neubert's formulation \cite{mom}, where 
resummation is achieved by using the running coupling-constant at the
vertices. The resummation integral is then a weighted average of the
coupling-constant at all scales:
\[
A_{res}=\int w(k)\alpha_s(k^2)d^4k
=\int_{-\infty}^\infty \rho(s)x^V(e^{s} Q^2)ds 
\]
where $w(k)$ is the Feynman integrand and 
$s=\ln\left(\frac{k^2}{Q^2}\right)$. 
The superscript $V$ stands for the V-scheme which is the most
convenient renormalization scheme for my purposes. 
While a specific scheme is used here in order to simplify the
formulae, it is important to understand that the 
above resummation integral is scheme-invariant \cite{mom}. 
The function $\rho(s)$ describes the
distribution of momentum of the exchanged gloun, and $x^V(e^{s}Q^2)$
describes the interaction strength as a function of the momentum. 
Using a 1-loop
formula for $x^V(e^{s}Q^2)$ I get:
\[
A_{res}=\int_{-\infty}^\infty \rho(s)
\left(\frac{x^V(Q^2)}{1\,+\,s\beta_0\,x^V(Q^2)}\right)ds  
\]
Clearly the integral includes contributions from an infinite set of diagrams. 
The exact distribution function (in the large $\beta_0$ approximation)
has been calculated for a few
observables, such as the vacuum-polarization D-function 
\cite{Broadhurst,BBB,mom} which I shall
use here as an example. A representative diagram is the following:
\begin{figure}[H]
  \begin{center}
\mbox{\kern-0.5cm
\epsfig{file=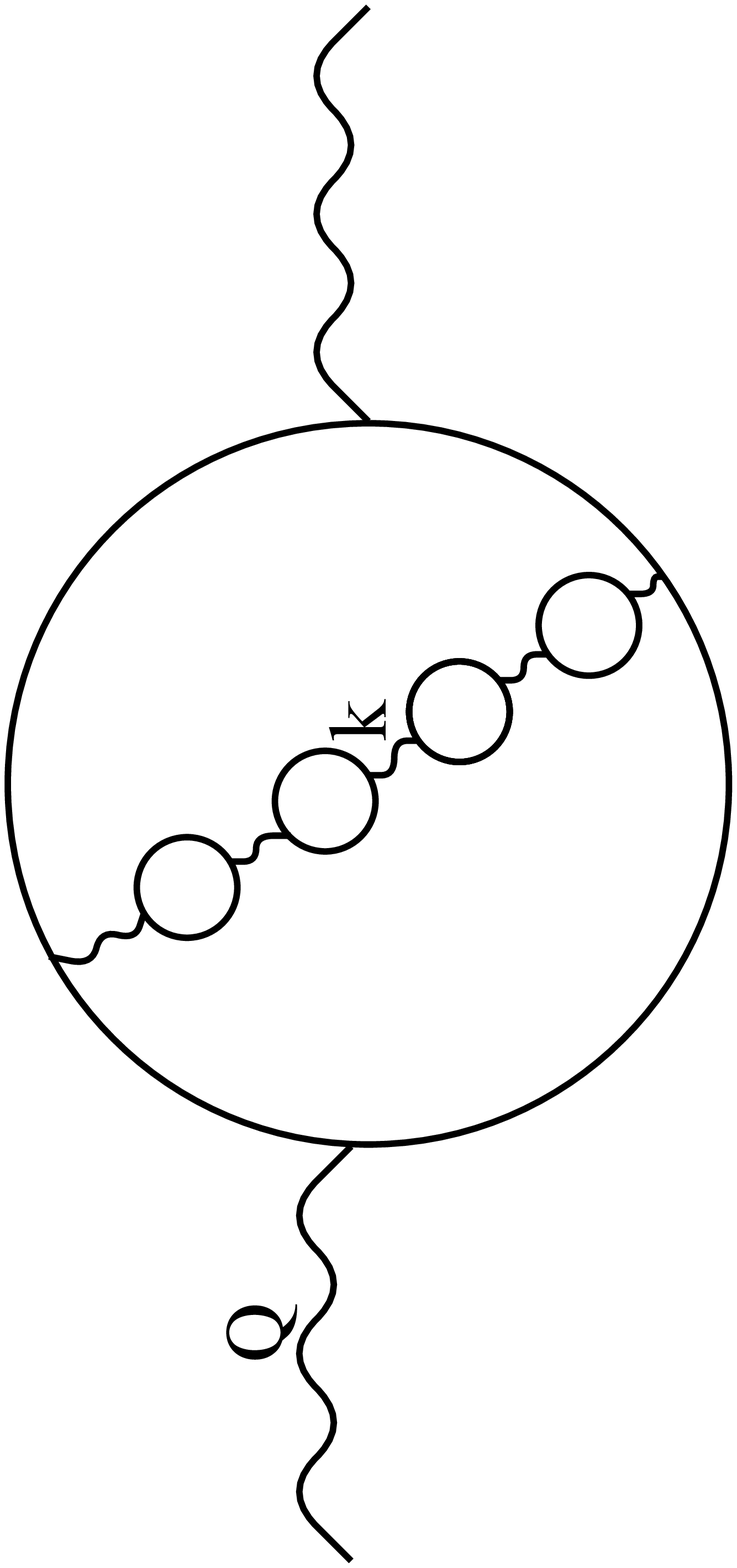,width=5.0truecm,angle=-90}}
\label{ren_chain}
  \end{center}
\end{figure}
In this particular diagram the exchanged gluon is dressed by fermion
loops.  However, gluonic
corrections that are related to the 1-loop running of the coupling are
resummed in the above technique just as well.
 
Of course, the resummation integral is not well defined, due to the
integration over the infra red pole in the 1-loop formula for the 
running coupling (Landau pole). This is how infra red renormalons
appear in this formulation. I will not deal here with the renormalon 
ambiguity which cannot be settled completely within perturbative QCD, 
but rather use the resummation integral to 
study the PA method. The general
methodology is to assume that only a first few coefficients in the 
perturbative series are known, construct a PA basing on this limited
information and then compare the PA with the exact all-order result.
We will see below that this comparison can be done also on the level of
the momentum distribution function, since PA's
can be interpreted as what one obtains by replacing the continuous
momentum distribution function with a particular discrete distribution.

It was found empirically that the momentum distribution function in
the large $\beta_0$ approximation is a non-negative function in many
physical examples \cite{mom,MDF}. This justifies a posteriori 
the probabilistic interpretation implied by the name `momentum distribution'.
If indeed $\rho(s) \geq 0$ for any $s$, then the
resummation integral defines a so-called Hamburger function \cite{Baker}:
\begin{equation}
f(z)\equiv A_{res}/\beta _0=\int_{-\infty }^\infty \rho (s)\frac
z{1+sz}ds=\int_{-\infty }^\infty \frac z{1+sz}d\phi (s).  
\label{f}
\end{equation}
where $z=\beta_0 x^V(Q^2)$ and $\phi(s)$ 
is the indefinite integral of $\rho(s)$.
The perturbative coefficients are moments of the distribution function: 
\[
f_i=\int_{-\infty }^\infty s^id\phi (s)  \label{f_i}
\]
for $i\geq 0$. 

There is a theorem \cite{Baker} that guarantees that for a Hamburger function, 
a $z[N-1/N]$ PA constructed from the partial-sum:
\[
z\sum_{i=0}^{2N-1}f_i(-z)^i
\]
can  be written as: 
\[
f(z)\,\sim \,z[N-1/N]\,=\,\sum_{i=1}^N\frac{ r_i z}
{1+q_i z}  
\]
with $q_i$  {\em real} and $r_i>0$ for $i=1,2,...N$.
 
Through this decomposition of the PA function, together with
eq.~(\ref{f}) one realizes that the PA corresponds to approximating 
the all-order continuous distribution function by a sum
of $N$ weighted $\delta$-functions:
\[
\rho _N(s)=\sum_{i=1}^N  r_i \delta (s- q_i)
\]
or, equivalently, its indefinite integral $\phi(s)$, by a piece-wise
constant function composed of $N$ steps:
\[
\phi _N(s)=\sum_{i=1}^N r_i \theta (s-  q_i).  
\]

Note that $\rho_N(s)$ is optimal (and unique, of course) since 
the equation for constructing the PA imply that $\rho_N(s)$ 
reproduces the first $2N$ moments of the distribution function, which
we know. Using a diagonal PA of a Hamburger function to identify
the optimal scales ($q_i$) and weights ($r_i$) is the basic idea
behind the method of Gaussian quadrature for numerical integration 
\cite{Baker}. 
 
In the Brodsky-Lepage-Mackenzie (BLM) approach \cite{BLM} one evaluates the 
coupling-constant at a scale that corresponds to the average momentum
of the exchanged gluon (the BLM scale). 
This is exactly equivalent to approximating the
distribution function by a single $\delta$-function located at its
center. This same effect can also be achieved simply by using a $x[0/1]$ PA
for the leading $\beta_0$ series \cite{Why}. 
In the method described above 
one uses an $x[N-1/N]$ PA of the leading $\beta_0$
series that corresponds to approximating the
momentum distribution function by a set of $N$ weighted $\delta$
functions. Therefore it can be viewed as a generalization of the BLM
method for higher-orders.

In the following, I illustrate the above ideas with 
the vacuum-polarization D-function. The all-order distribution
function $\rho(s)$ \cite{mom} is plotted in fig.~6 as a continuous
line. We see that
there are contributions from both UV scales (positive $s$) and IR
scales (negative $s$). 
The different symbols correspond to the
locations ($q_i$) and weights ($r_i$) of the diagonal PA's poles. 
For the $x[0/1]$
PA, there is one $\delta$-function at the BLM scale. For the $x[1/2]$
PA, there are two $\delta$-functions, and so on. 
In Fig.~7 we see how using $x[N-/N]$ PA corresponds to approximating 
the integral distribution function $\phi(s)$ 
by a piecewise constant function, composed of $N$ steps.

To conclude, we saw that diagonal PA's can be used to resum certain all-order
effects that are related to the running of the coupling constant, and
thus provide a systematic method for obtaining reliable scale-invariant
predictions. There is a rigorous relation between diagonal 
PA and the momentum distribution of virtual gluons. I stress that
this result holds only for a
single gluon exchange, i.e. within the large $\beta_0$
approximation. The way to go beyond this approximation is still unclear.
Nevertheless, from the results presented here for the Bjorken Sum-Rule,
it is clear that PA's are an important tool for QCD phenomenology.  

\vspace{.7cm}
\begin{flushleft}
{\large\bf Acknowledgments}
\end{flushleft}
         
I thank Marek Karliner, Stan Brodsky, John Ellis and Mark Samuel for
a fruitful cooperation and very useful discussions. 
The research was supported in part by the Israel
Science Foundation administered by the Israel Academy of Sciences and
Humanities, and by a Grant from the G.I.F., the German-Israeli
Foundation for Scientific Research and Development.

\begin{figure}[H]
  \begin{center}
\mbox{\kern-0.5cm
\epsfig{file=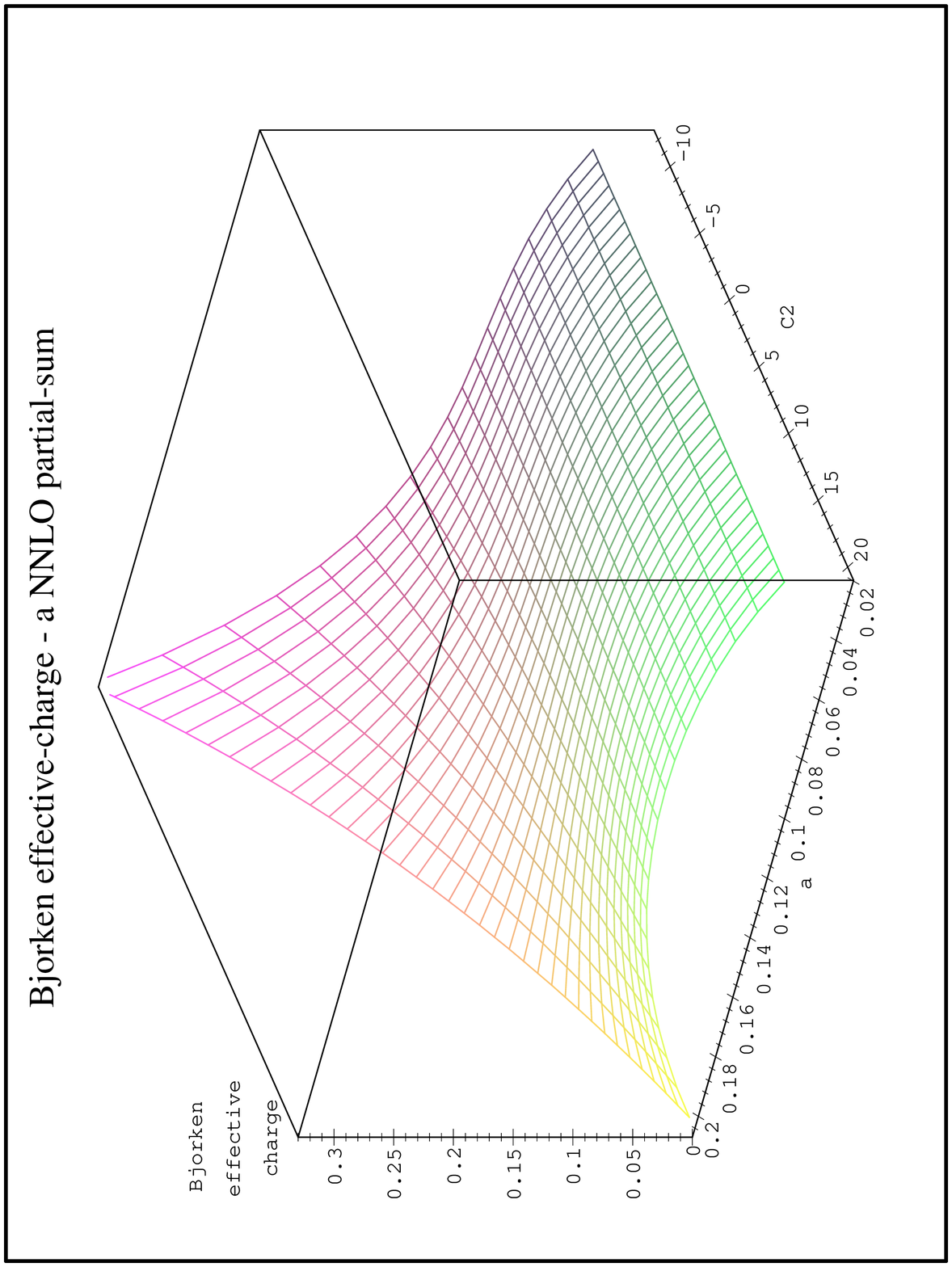,width=11.0truecm,angle=0}}
\label{f1}
  \end{center}
{\Large
Fig. 1}
\end{figure}
\newpage

\begin{figure}[H]
  \begin{center}
\mbox{\kern-0.5cm
\epsfig{file=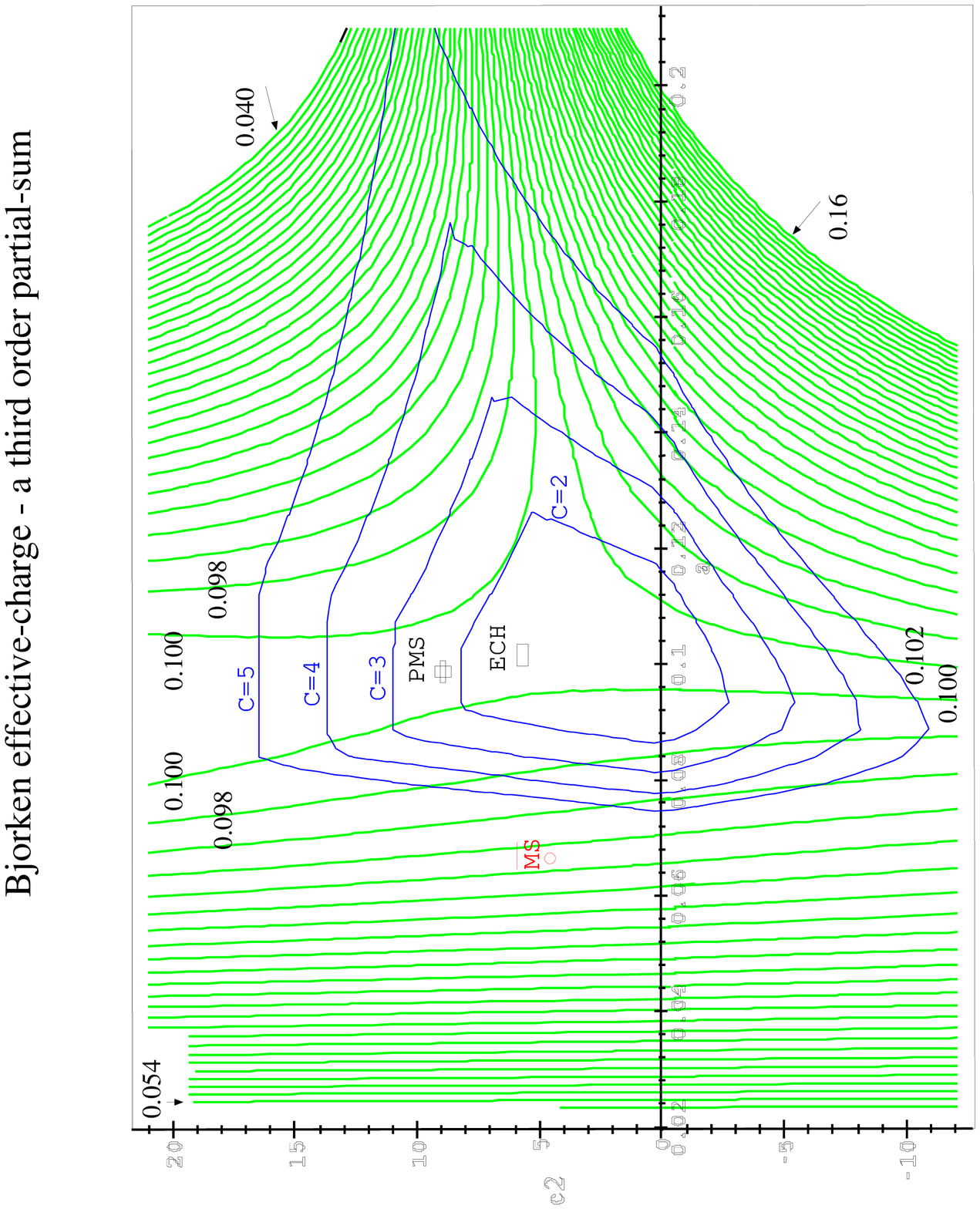,width=11.0truecm,angle=180}}
\label{f2}
  \end{center}
{\Large
Fig. 2}
\end{figure}
\newpage

\begin{figure}[H]
  \begin{center}
\mbox{\kern-0.5cm
\epsfig{file=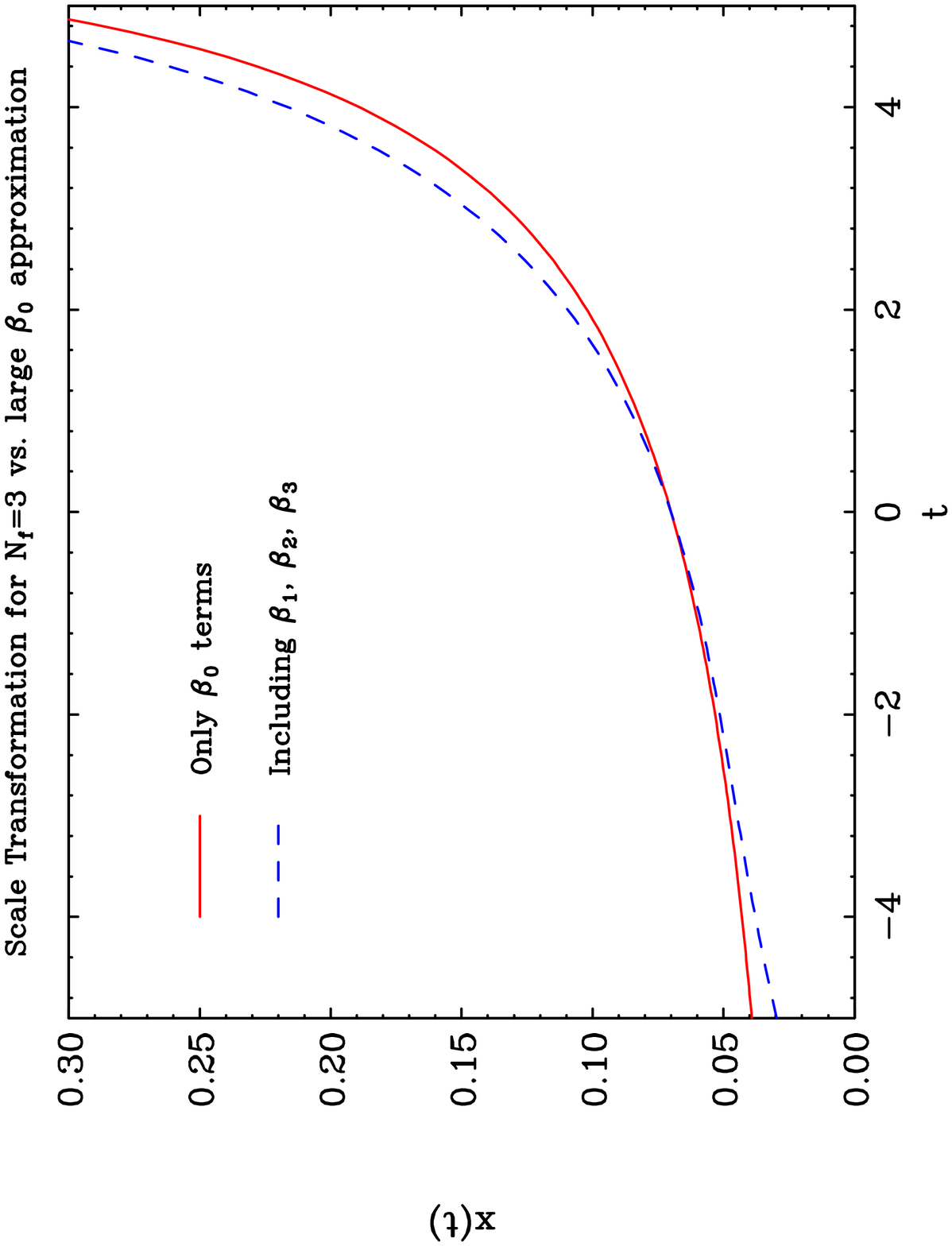,width=11.0truecm,angle=180}}
\label{f3}
  \end{center}
{\Large
Fig. 3}
\end{figure}
\newpage

\begin{figure}[H]
  \begin{center}
\mbox{\kern-0.5cm
\epsfig{file=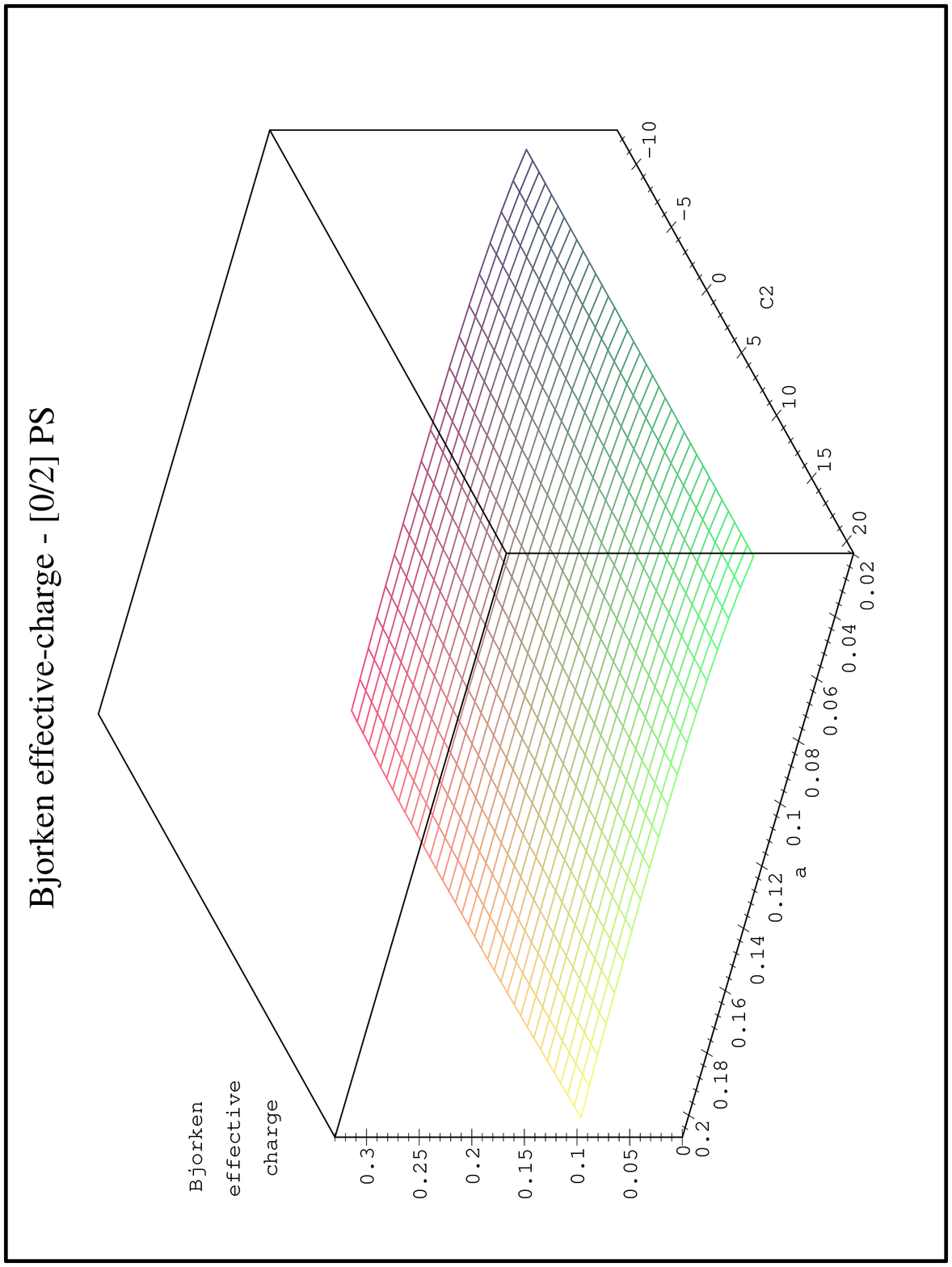,width=11.0truecm,angle=0}}
\label{f4}
  \end{center}
{\Large
Fig. 4}
\end{figure}
\newpage

\begin{figure}[H]
  \begin{center}
\mbox{\kern-0.5cm
\epsfig{file=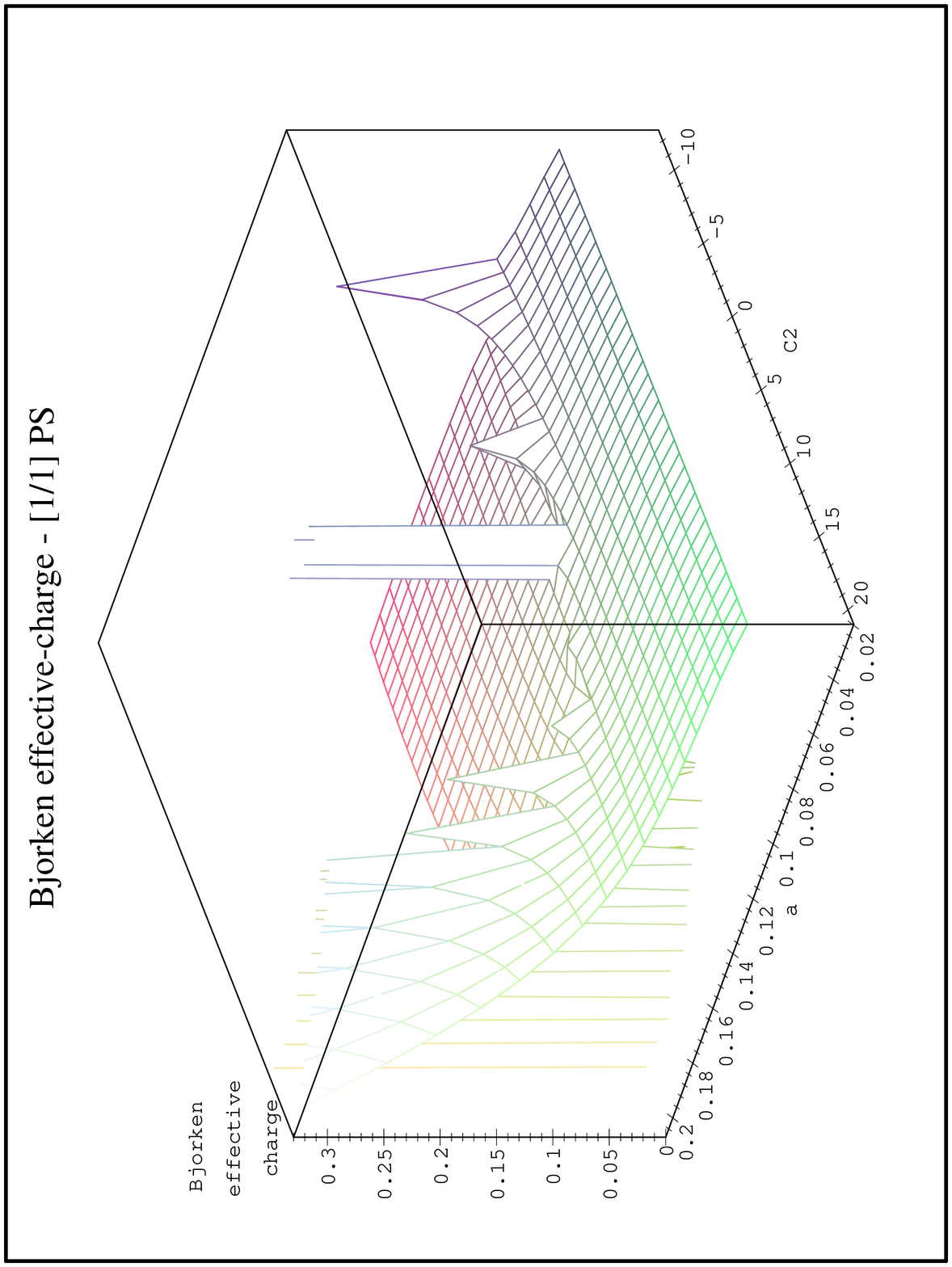,width=11.0true
  cm,angle=0}}
\label{f5}
  \end{center}
{\Large
Fig. 5}
\end{figure}
\newpage

\begin{figure}[H]
  \begin{center}
\mbox{\kern-0.5cm
\epsfig{file=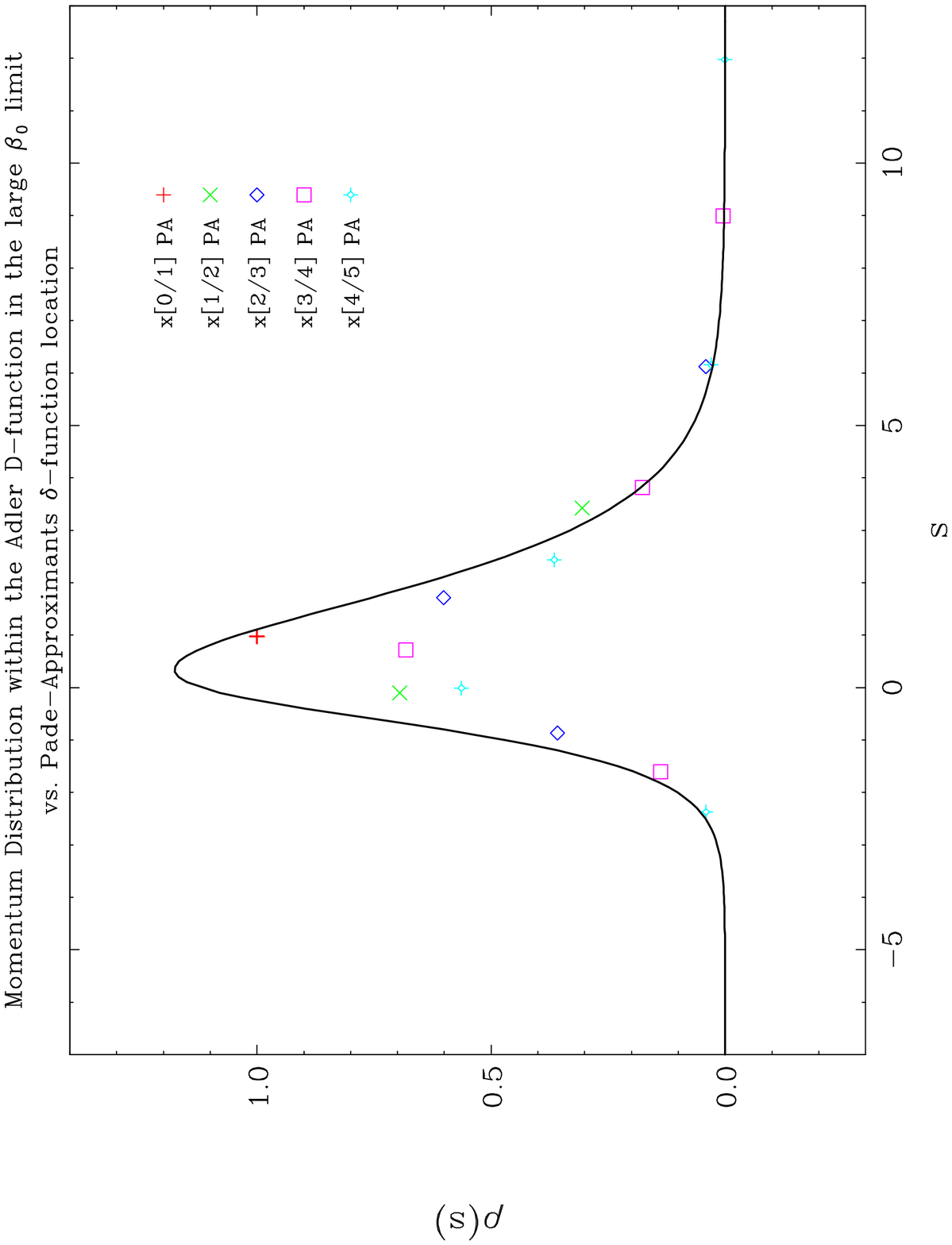,width=11.0truecm,angle=180}}
\label{f6}
  \end{center}
{\Large
Fig. 6}
\end{figure}
\newpage

\begin{figure}[H]
  \begin{center}
\mbox{\kern-0.5cm
\epsfig{file=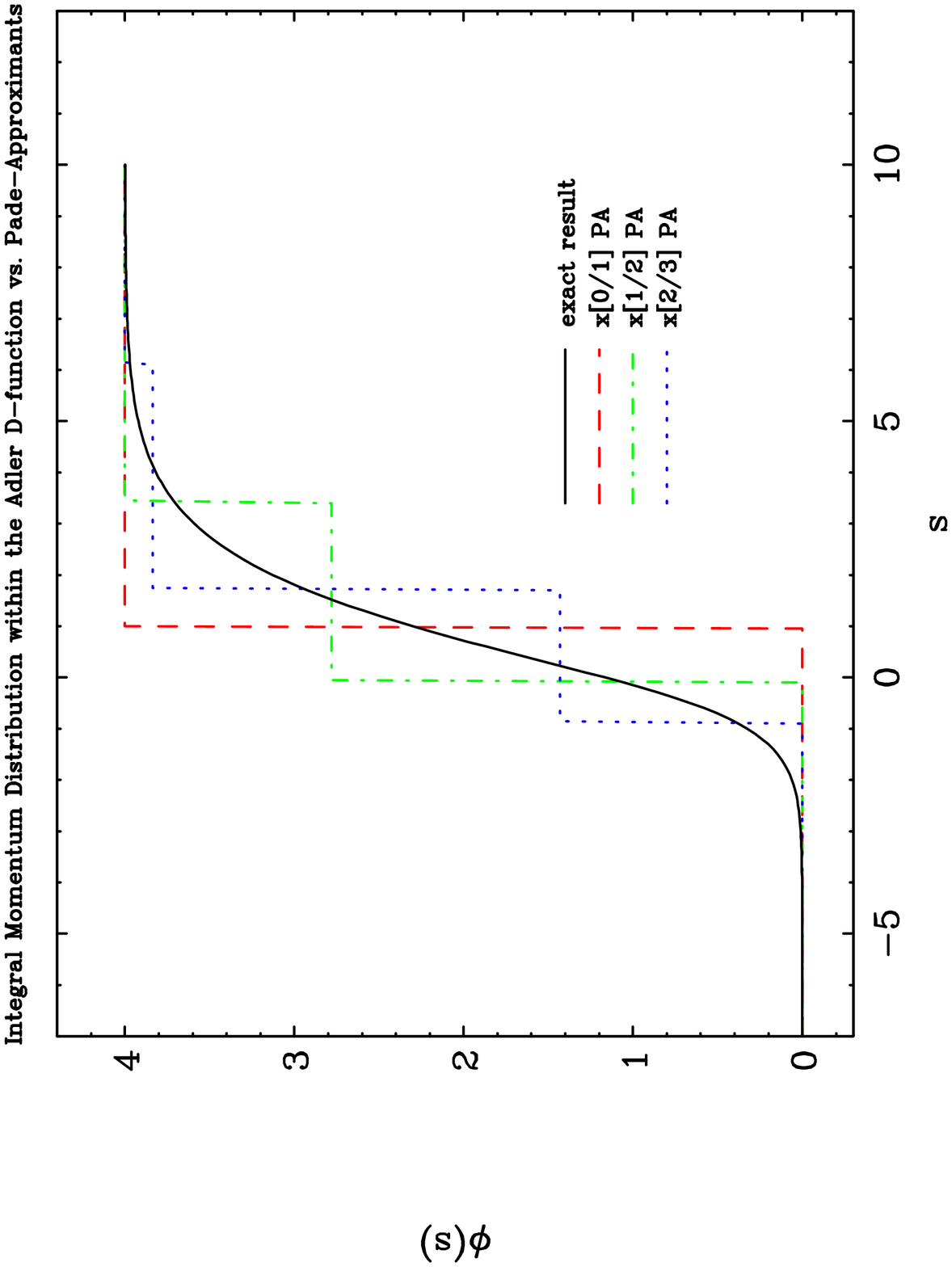,width=11.0truecm,angle=180}}
\label{f7}
  \end{center}
{\Large
Fig. 7}
\end{figure}


\begin{thebibliography}{99}

\bibitem{Fischer}
J. Fischer,
{\em On the Role of Power Expansions in Quantum Field Theory},
hep-ph/9704351.

\bibitem{padeworks}  M.A. Samuel, G. Li and E. Steinfelds, {\it Phys. Rev.} 
{\bf D48}(1993)869 and {\it Phys. Lett.} {\bf B323}(1994)188; M.A. Samuel
and G. Li, {\it Int. J. Th. Phys.} {\bf 33}(1994)1461 and {\it Phys. Lett.} 
{\bf B331}(1994)114.

\bibitem{PA_QCD}  M.A. Samuel, J. Ellis and M. Karliner, 
{\sl Phys. Rev. Lett.} {\bf 74}(1995)4380;
J. Ellis, E. Gardi, M. Karliner and M.A. Samuel, 
{\it Phys. Lett.} {\bf B366}(1996)268
and
{\it Phys.~Rev.} {\bf D54}(1996)6986;
I.~Jack, D.R.T.~Jones and M.A.~Samuel,
{\em Asymptotic Pad\'e Approximants and the SQCD $\beta$ Function},
hep-ph/9706249.

\bibitem{Why}  
E. Gardi, 
{\em Phys. Rev.} {\bf D56}(1997)68.

\bibitem{MDF}
S.J. Brodsky, J. Ellis, E. Gardi, M. Karliner and M.A. Samuel, 
{\em Pad\'e Approximants, Optimal Renormalization Scales, and Momentum
  Flow in Feynman Diagrams}, 
hep-ph/9706467, CERN-TH-97/126. Accepted for publication in 
{\it Phys. Rev.} {\bf D}.

\bibitem{Baker}  George A. Baker, Jr. and Peter Graves-Morris, \thinspace
{\em Gian-Carlo Rota Encyclopedia of Mathematics and its Applications}
(Addison-Wesley, 1981), Vol. 13 and 14.


\bibitem{ECH}  G. Grunberg, {\it Phys. Rev.} {\bf D29}(1984)2315.

\bibitem{PMS}  P.M. Stevenson, {\it Phys. Rev.} {\bf D23}(1981)2916.

\bibitem{BLM}  S.J. Brodsky, G.P. Lepage and P.M. Mackenzie, {\em Phys.
Rev.\ } {\bf D28}(1983)228;



\bibitem{mom}  M. Neubert, {\it Phys. Rev.} {\bf D51}(1995)5924.

\bibitem{BB}
M. Beneke and V.M. Braun,
{\em Phys. Lett.} {\bf B348}(1995)513.

\bibitem{LTM}  C.N. Lovett-Turner and C.J. Maxwell, 
{\em Nucl. Phys.} {\bf B432}(1994)147
and 
{\em Nucl. Phys.\ } {\bf B452}(1995)188.

\bibitem{Broadhurst}  D.J. Broadhurst, {\it Z. Phys} {\bf C58}(1993)339.

\bibitem{BBB}  M. Beneke, {\it Phys. Lett.} {\bf B307}(1993)154; 
{\em Nucl. Phys.} {\bf B405}(1993)424.





\end{thebibliography}
\end{document}